\documentclass[12pt]{iopart}

\usepackage{graphicx}
\usepackage{color}

\begin{document}

\title{Bound states for the quantum dipole moment in two dimensions}

\author{Paolo Amore\dag \ and Francisco M Fern\'andez
\footnote[2]{Corresponding author}}

\address{\dag\ Facultad de Ciencias, Universidad de Colima, Bernal D\'iaz del
Castillo 340, Colima, Colima, Mexico}\ead{paolo.amore@gmail.com}

\address{\ddag\ INIFTA (UNLP,CCT La Plata-CONICET), Divisi\'on Qu\'imica Te\'orica,
Blvd. 113 S/N,  Sucursal 4, Casilla de Correo 16, 1900 La Plata,
Argentina}\ead{fernande@quimica.unlp.edu.ar}

\maketitle

\begin{abstract}
We calculate accurate eigenvalues and eigenfunctions of the Schr\"odinger
equation for a two-dimensional quantum dipole. This model proved useful for
the study of elastic effects of a single edge dislocation. We show that the
Rayleigh-Ritz variational method with a basis set of Slater-type functions
is considerably more efficient than the same approach with the basis set of
point-spectrum eigenfunctions of the two-dimensional hydrogen atom used in
earlier calculations.
\end{abstract}

\section{Introduction}

In a recent paper Dasbiswas et al\cite{DGYD10} discussed the bound-state
spectrum for a straight edge dislocation oriented along the $z$ axis. Within
the continuum model, the authors reduced the problem to the two-dimensional
Schr\"{o}dinger equation for a quantum dipole. They obtained very accurate
eigenvalues and eigenfunctions by means of a discretization of the $x-y$
space. Since this real-space diagonalization method (RSDM) is not so
practical for highly excited states, those authors also carried out a
Rayleigh-Ritz (RR) variational calculation with the basis set of
eigenfunctions of the two-dimensional Coulomb problem\cite{YGWC91}.

The variational RR eigenvalues are known to approach the exact ones from
above. However, in the present case the ground-state energy $\epsilon
_{1}^{RR}=-0.0970$ calculated with as many as $400$ basis functions exhibits
a considerable discrepancy with respect to the same eigenvalue obtained by
RSDM $\epsilon _{1}^{RSDM}=-1.39$. Later on, Amore\cite{A12} carried out a
more accurate calculation with $3600$ hydrogen eigenfunctions and obtained $%
\epsilon _{1}^{RR}=-0.128$ as well as $\epsilon _{1}^{RRS}=-0.132$ by
fitting and extrapolating the outcome of a Shanks transformation. The
authors of both articles resorted to a variational parameter (decaying
parameter\cite{DGYD10} or length scale\cite{A12}) that considerably improves
the result. The remaining disagreement between RR and RSDM is probably due
to the well known fact that the basis set used in those calculations is not
complete because it does not include the continuous spectrum\cite{YGWC91}
(see, for example Ref.~\cite{CCCGA12} and the references therein). For this
reason the RR calculation proposed by those authors\cite{DGYD10,A12} is
expected to have a limited accuracy no matter how large the dimension of the
basis set of discrete states. It is surprising, however, that the lack of
the continuous wavefunctions appears to be more noticeable for the ground
state. In addition to it, at first sight it seems that there is a better
agreement for the odd states\cite{DGYD10}.

The purpose of this paper is to carry out a RR calculation with a
nonorthogonal basis set of square-integrable functions that in principle
does not require the continuous spectrum. In section~\ref{sec:RR} we
describe the RR variational method with such an improved basis set. In
section~\ref{sec:results} we compare present results with those obtained
earlier by Dasbiswas et al\cite{DGYD10} and Amore\cite{A12}. Finally, in
section~\ref{sec:conclusions} we summarize the main results and draw
conclusions.

\section{Rayleigh-Ritz variational method}

\label{sec:RR}

The linearized model for the Ginzburg-Landau theory leads to the
Schr\"{o}dinger equation

\begin{equation}
-\frac{\hbar ^{2}}{2m}\nabla ^{2}\psi +p\frac{\cos \theta }{r}\psi =E\psi
\label{eq:Schro}
\end{equation}
where $r=\sqrt{x^{2}+y^{2}}$, $0\leq \theta =\arctan (y/x)<2\pi $, and $p$
is the strength of the dipole potential\cite{DGYD10}. Choosing the units of
length $\hbar ^{2}/(2mp)$ and energy $2mp^{2}/\hbar ^{2}$ we obtain the
dimensionless eigenvalue equation
\begin{equation}
-\nabla ^{2}\psi +\frac{\cos \theta }{r}\psi =\epsilon \psi
\label{eq:Schro_dim}
\end{equation}
where $\epsilon =\hbar ^{2}E/(2mp^{2})$. Since the potential $V(r,\theta
)=\cos \theta /r$ is invariant under reflection about the $x$ axis $%
V(r,-\theta )=V(r,\theta )$, then the wavefunctions are either even $\psi
(r,-\theta )=\psi (r,\theta )$ or odd $\psi (r,-\theta )=-\psi (r,\theta )$
under such coordinate transformation.

Dasbiswas et al\cite{DGYD10} and Amore\cite{A12} resorted to the RR
variational method with the basis set of eigenfunctions of the planar
hydrogen atom:
\begin{equation}
\psi _{n,l}^{H}(r,\theta )=\frac{1}{\sqrt{\pi }}R_{n,l}(r)\times \left\{
\begin{array}{c}
\cos (l\theta ),\,1\leq l\leq n \\
\frac{1}{\sqrt{2}},l=0 \\
\sin (l\theta ),\,-n\leq l\leq -1
\end{array}
\right.  \label{eq:Hyd_basis}
\end{equation}
where $n=1,2,\ldots $ and $R_{n,l}(r)$ is the normalized solution to the
radial equation\cite{YGWC91}. However this basis set is incomplete if one
does not include the eigenfunctions for the continuous spectrum (see, for
example,\cite{CCCGA12} and references therein).

If, on the other hand, the chosen basis set of square-integrable functions
does not require the continuous spectrum to be complete then one expects the
accuracy of the RR variational results to be determined only by the basis
dimension. Here we propose the nonorthogonal set of functions
\begin{equation}
\{\phi _{j}^{e},\,j=1,2,\ldots \}=\left\{ e^{-\alpha r},r^{i+1}\cos
^{j}\theta e^{-\alpha r},\,i,j=0,1,\ldots \right\}  \label{eq:basis_e}
\end{equation}
for the even states and
\begin{equation}
\{\phi _{j}^{o},\,j=1,2,\ldots \}=\left\{ r^{i+1}\sin \theta \cos ^{j}\theta
e^{-\alpha r},\,i,j=0,1,\ldots \right\}  \label{eq:basis_o}
\end{equation}
for the odd ones, where $\alpha >0$ is a variational parameter. This basis
set resembles the Slater orbitals commonly used in quantum chemistry
calculations of atomic and molecular electronic structure\cite{CCCGA12}.

The RR method with the variational ansatz
\begin{equation}
\psi =\sum_{j=m}^{N}c_{m}\phi _{m}  \label{eq;wave_exp}
\end{equation}
leads to the generalized eigenvalue problem
\begin{equation}
\mathbf{HC}=\epsilon \mathbf{SC}
\end{equation}
where $H_{ij}=\left\langle \phi _{i}\right| \hat{H}\left| \phi
_{j}\right\rangle $ and $S_{ij}=\left\langle \phi _{i}\right| \left. \phi
_{j}\right\rangle $. Note that it is possible to obtain explicit expressions
for both kinds of matrix elements in terms of the variational parameter%
\textbf{\ $\alpha $; }even more important is the fact that, with the help of
a computer algebra software, like Mathematica, we can obtain the inverse of $%
\mathbf{S}$ explicitly for all the cases considered in the present paper%
\textbf{.} In this way the inversion does not introduce any round-off errors
and the original generalized eigenvalue problem is converted to the ordinary
eigenvalue problem
\begin{equation}
\mathbf{S}^{-1}\mathbf{HC}=\epsilon \mathbf{C}
\end{equation}
for the nonsymmetric matrix $\mathbf{S}^{-1}\mathbf{H}$.

\section{Results}

\label{sec:results}

\begin{figure}[h]
~\bigskip\bigskip
\par
\begin{center}
\includegraphics[width=9cm]{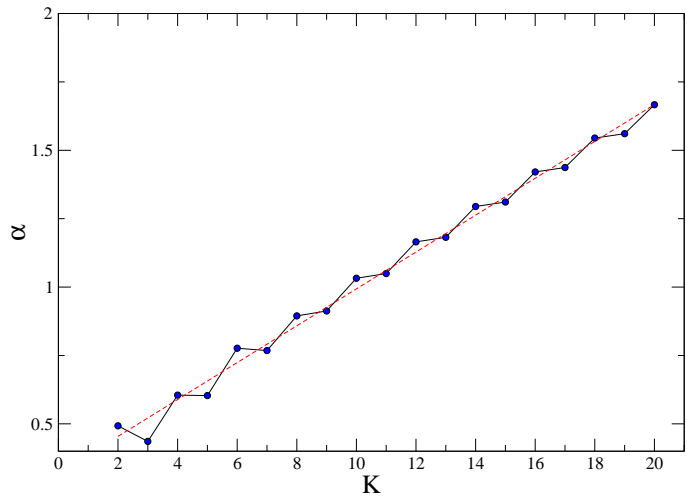} \bigskip
\end{center}
\caption{Optimal value of $\alpha$ for the ground state as a
function of $K$} \label{Fig:alpha(K)}
\end{figure}

\begin{figure}[h]
~\bigskip\bigskip
\par
\begin{center}
\includegraphics[width=9cm]{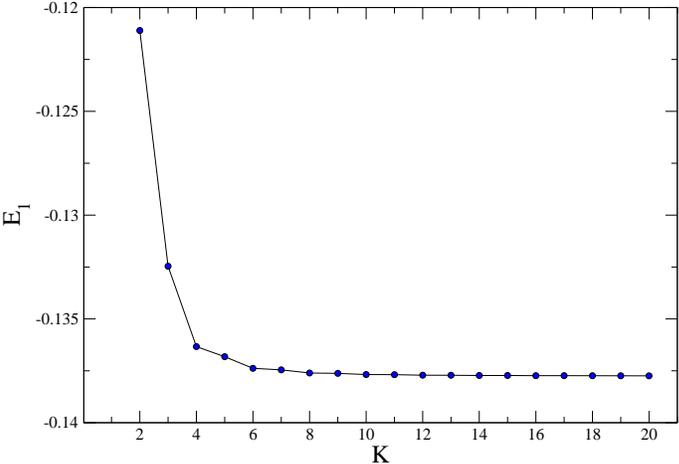} \bigskip
\end{center}
\caption{Convergence of $\epsilon_1$ in terms of $K$ for the optimal value
of $\alpha$ }
\label{Fig:E1(K)}
\end{figure}

For brevity we write both the exact and approximate RR eigenvalues and
eigenfunctions as $\epsilon _{1}<\epsilon _{2}<\ldots $ and $\psi _{1},\psi
_{2},\ldots $, respectively. We express the rate of convergence of the RR
results in terms of the largest parameter $K=i+j$ in the wavefunction
expansion (\ref{eq;wave_exp}), where $i$ and $j$ are the exponents of $r$
and $\cos \theta $ in either equation (\ref{eq:basis_e}) or (\ref{eq:basis_o}%
). Thus, for a given value of $K$ there are $N=(K^{2}+K+2)/2$ basis
functions of either even or odd symmetry. The optimal value of the nonlinear
variational parameter $\alpha $ depends on both the chosen eigenvalue and
the number of terms $N$ in the wavefunction expansion (\ref{eq;wave_exp}).
For example,  Fig.~\ref{Fig:alpha(K)} shows that for the ground state $%
\alpha $ increases with $K$ oscillating about a straight line. Fig.~\ref
{Fig:E1(K)} shows the RR eigenvalue $\epsilon _{1}$ for a range of values of
$K$. The rate of convergence is remarkably greater than the one for the
Coulomb basis set\cite{A12}.

Table~\ref{tab:energies} shows the RR eigenvalues obtained with the
nonorthogonal basis sets (\ref{eq:basis_e}) and (\ref{eq:basis_o}) ($%
\epsilon _{n}^{NB}$) on the one side and with the basis sets of even and odd
Coulomb functions (\ref{eq:Hyd_basis}) ($\epsilon _{n}^{CB}$) on the other.
We appreciate the following facts: the accuracy of present nonlinear basis
set is always greater ($\epsilon _{n}^{NB}\leq \epsilon _{n}^{CB}$) in spite
of the fact that the RR calculations have been carried out with with $N=211$
nonorthogonal basis functions ($K=20)$ and $N=3600$ Coulomb functions\cite
{A12}. The discrepancy is greater for the lowest states and, it is less
noticeable for the odd ones.  We are presently unable to provide a rigorous
proof for the last two facts; however, we may conjecture that the omitted
continuous spectrum is not so relevant in those cases where there is
agreement between the NB and CB variational results\textbf{.} The number of
digits in the entries of this table is dictated by comparison purposes and
does not reflect the estimated accuracy of the calculation.

Both the analytical calculation of the matrix elements and the analytical
inversion of the matrix\textbf{\ }$\mathbf{S}$ are time consuming but we do
them only once for all the states. On the other hand, the optimization of
the variational parameter\textbf{\ }$\alpha $ for every state is a time
consuming calculation that we should repeat several times. Earlier and
present calculations suggest that the RR variational method is less
efficient for the ground state. For this reason we have attempted
variational calculations with considerably greater basis sets only for this
state. For example, we have obtained\textbf{\ $\epsilon
_{1}^{NB\,\,even}=-0.13774677227$} with\ $N=466$ NB functions\ ($K=30\mathbf{%
)}$ and $\epsilon ^{NB\,even}=-0.13774778205$ with\ $N=821$ ones ($K=40$).
These results suggest that the first 6 digits remain stable and,
consequently, that the RR calculation with the Slater-type basis functions
does not appear to approach the RSDM results any more closely. However, it
is worth noticing that present RR eigenvalues agree with the RSDM ones
within the $2\%$ error estimated by Dasbiswas et al\cite{DGYD10}.

By means of the approximate ground-state wavefunction $\psi _{1}(x,y)$ in
terms of the Slater-like orbitals we have also calculated the effective
dimensionless coupling constant\cite{DGYD10}
\begin{equation}
g=\int dxdy|\psi _{1}(x,y)|^{4}
\end{equation}
Dasbiswas et al\cite{DGYD10} obtained $g=0.017$ by means of a simple
variational function constructed from the first three elements of the even
nonorthogonal basis set (\ref{eq:basis_e}): $\{e^{-\alpha r},re^{-\alpha
r},r\cos \theta e^{-\alpha r}\}$ and $g=0.0194$ by means of the RSDM. Amore%
\cite{A12} obtained $g=0.017$ by means of a reduced RR basis set of Coulomb
functions ($N=345$). Note that just $N=3$ NB functions yield the same result
as $N=345$ Coulomb ones. The RR method with $N=211$ functions ($K=20$) of
the basis set (\ref{eq:basis_e}) yields $g=0.0193$ that is quite close to
the RSDM result.

\begin{figure}[h]
~\bigskip\bigskip
\par
\begin{center}
\includegraphics[width=5cm]{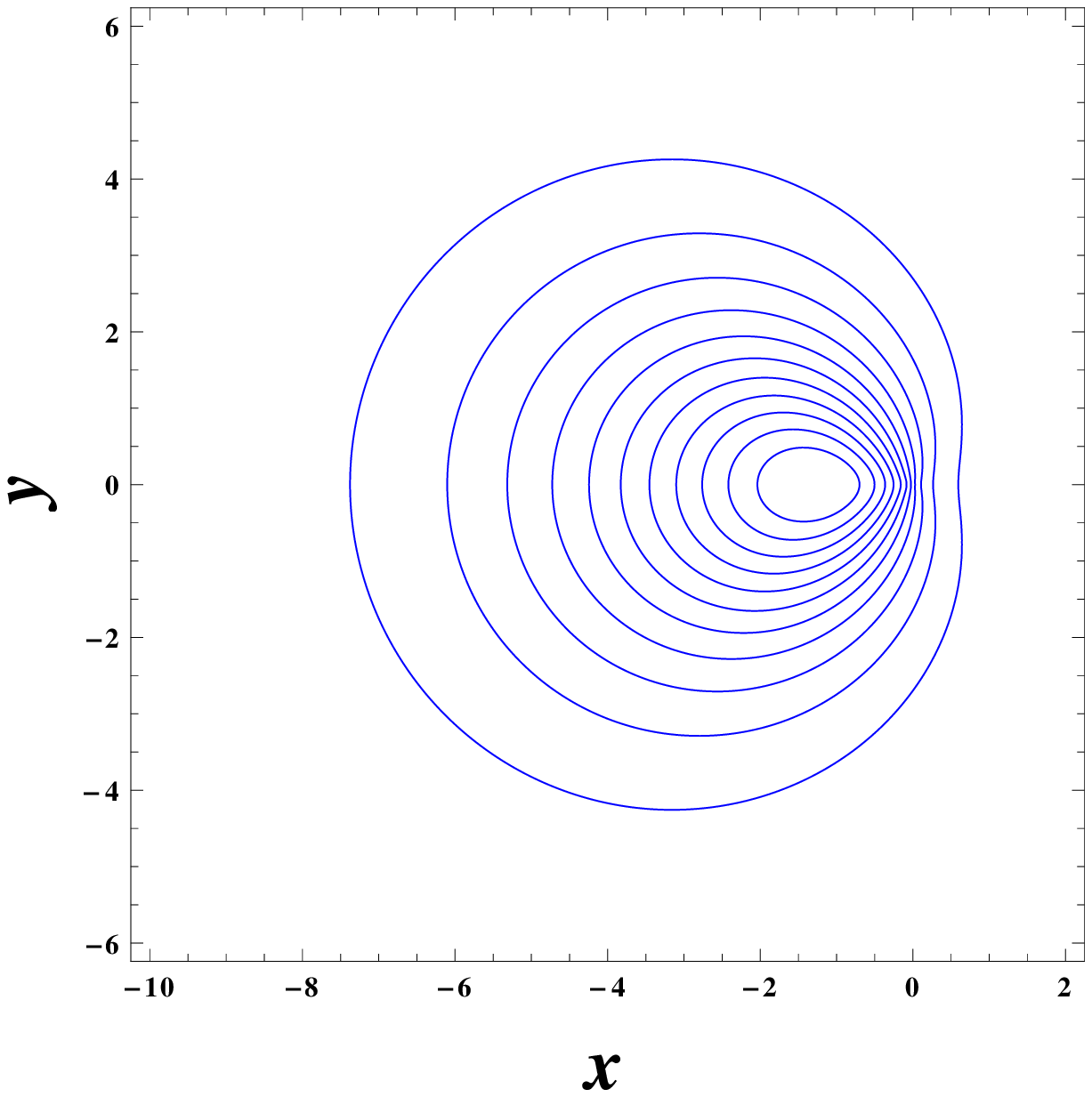} \bigskip %
\includegraphics[width=6cm]{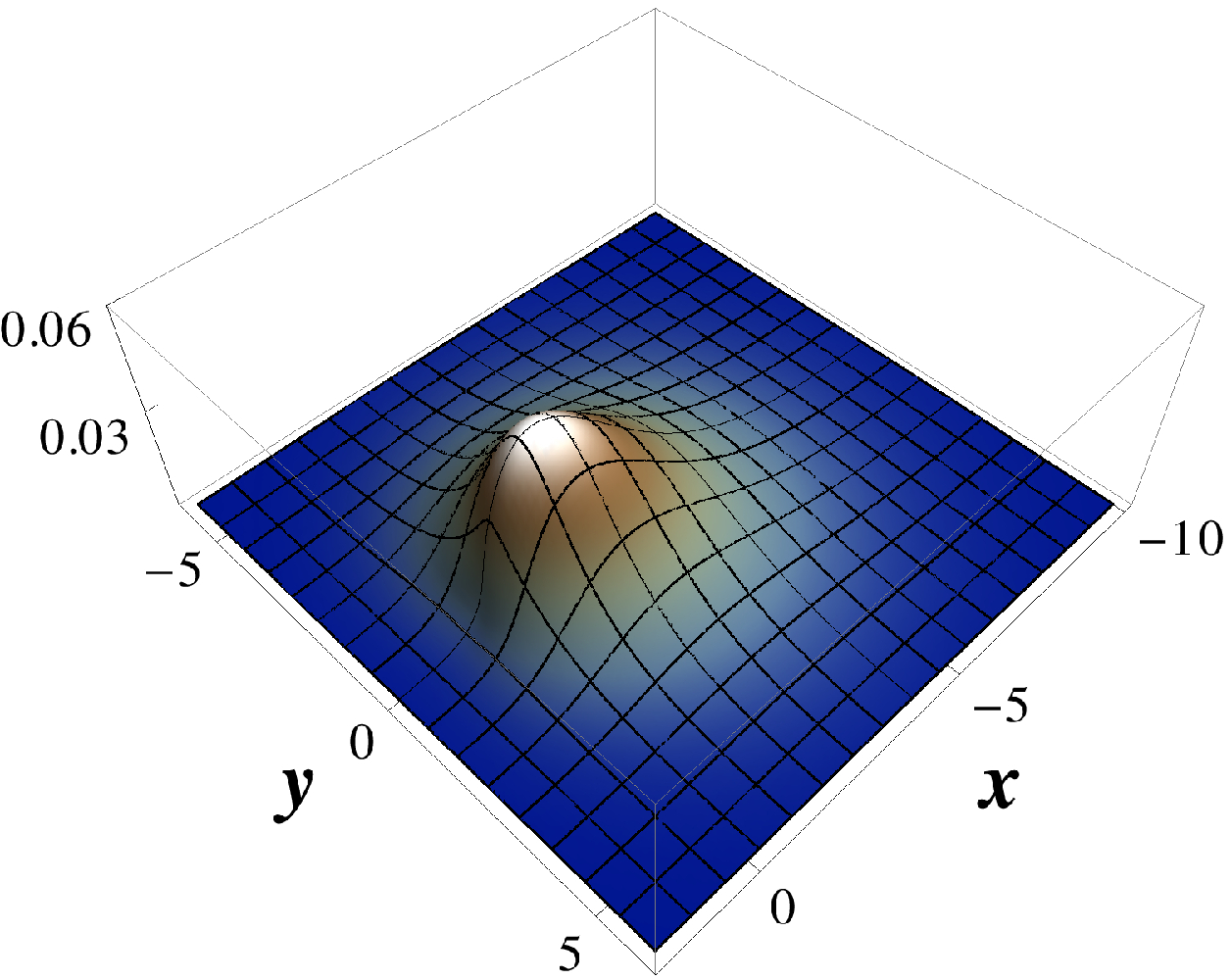}
\end{center}
\caption{Contour and 3D plot for $|\psi_1 |^2$}
\label{Fig:density_1_even}
\end{figure}

\begin{figure}[h]
~\bigskip\bigskip
\par
\begin{center}
\includegraphics[width=4cm]{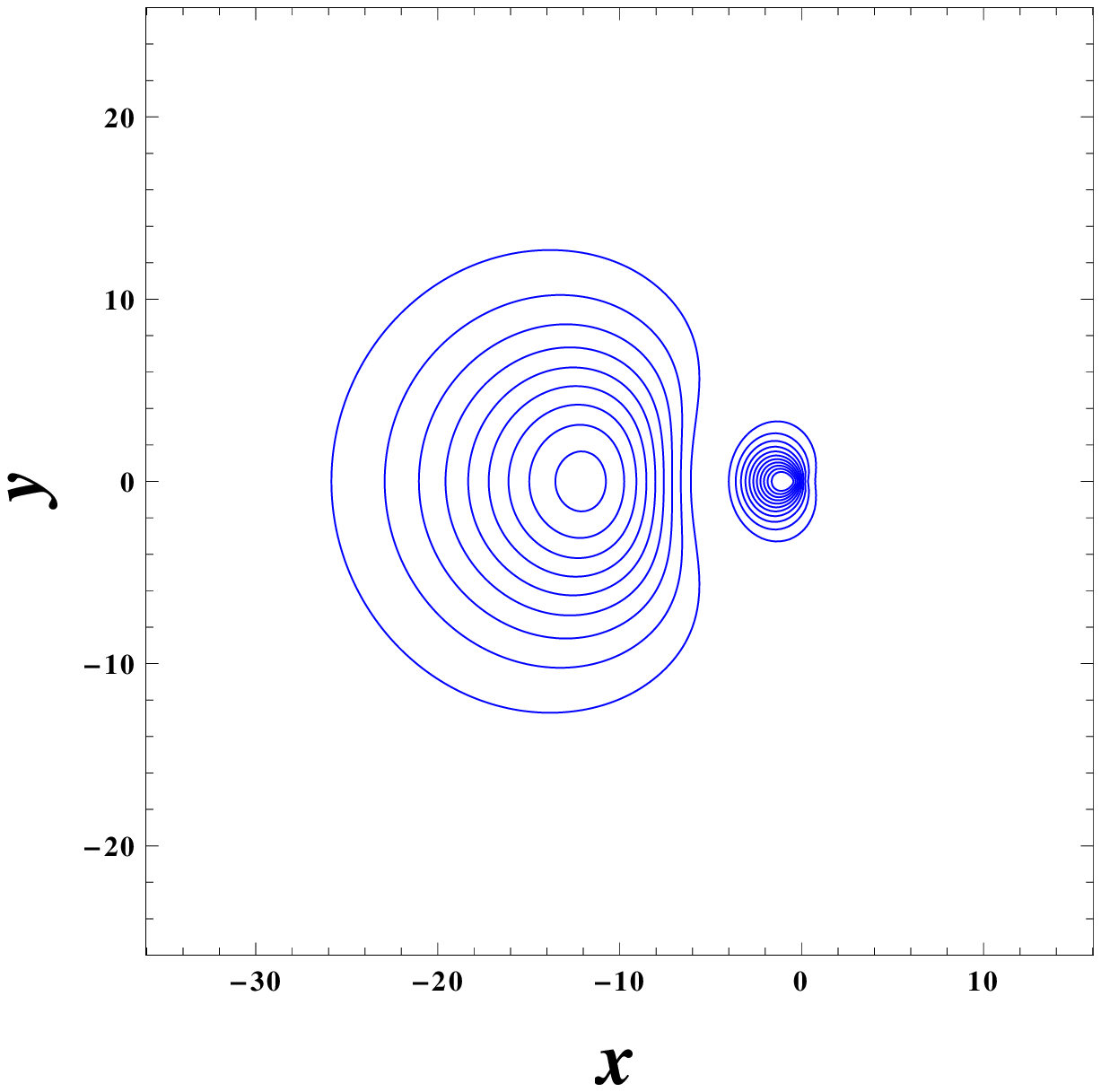} %
\includegraphics[width=4cm]{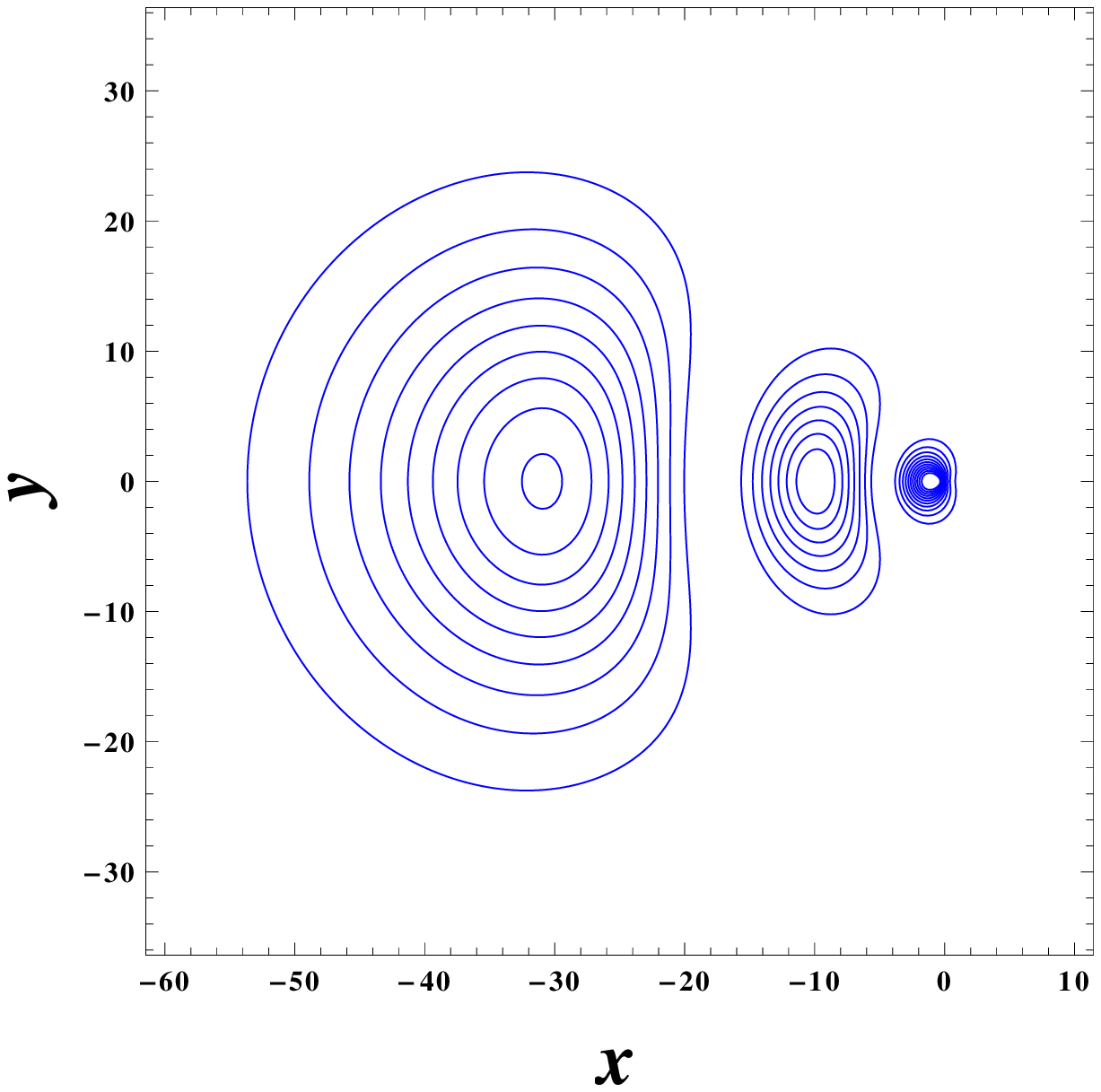}\\[0pt]
\includegraphics[width=4cm]{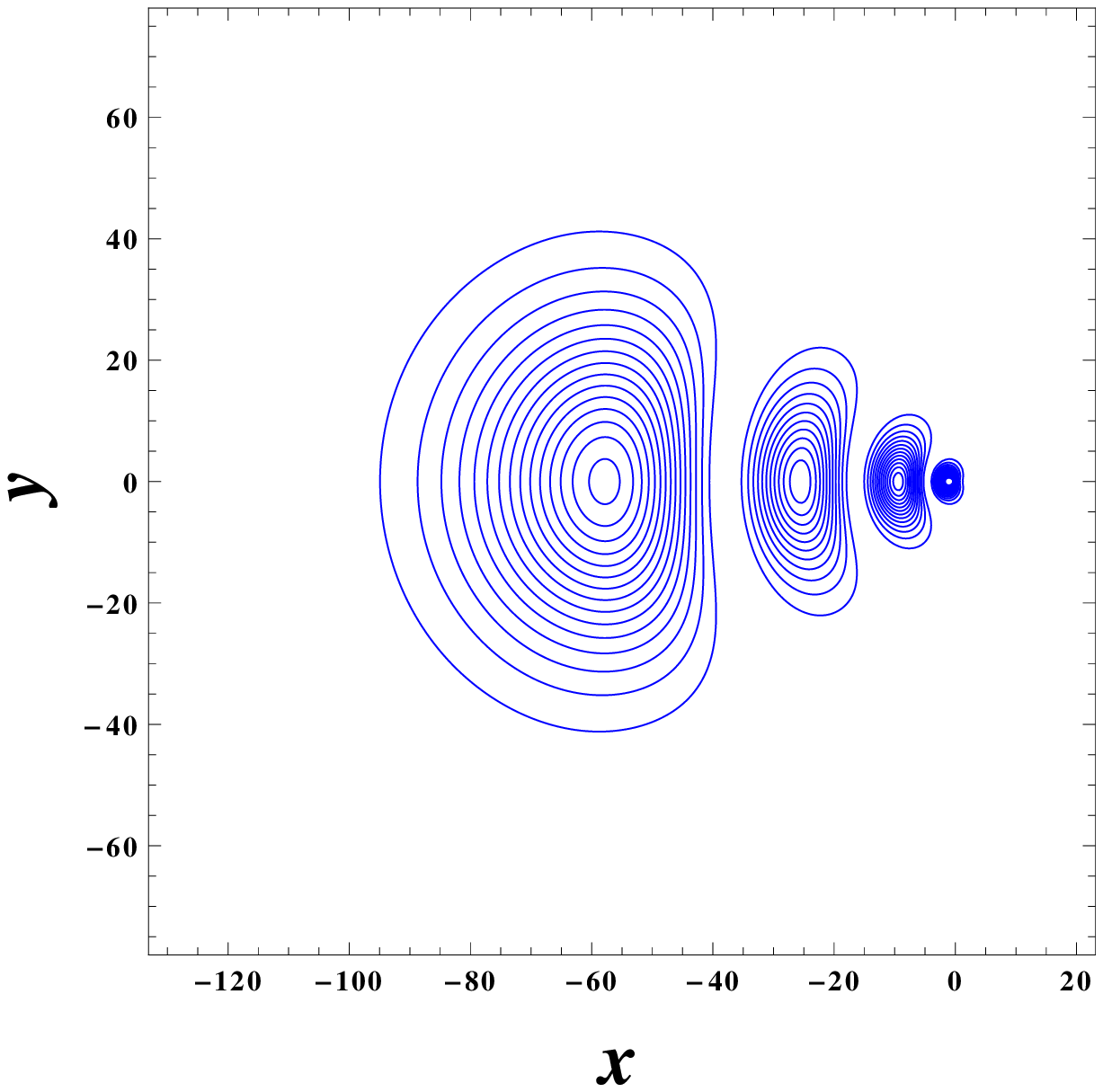} %
\includegraphics[width=4cm]{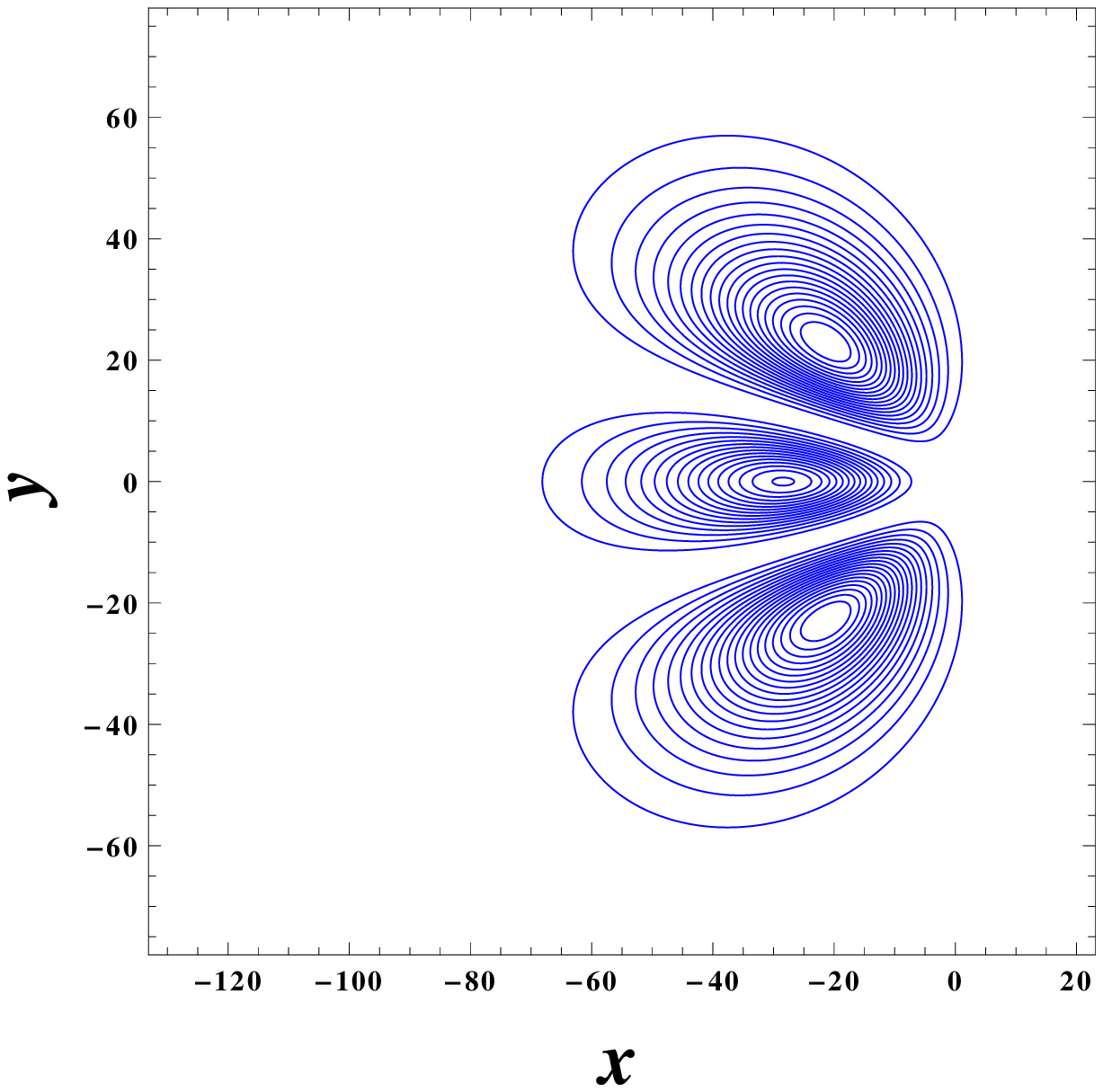}\bigskip
\end{center}
\caption{Contour plots for the squares of the second, third,
fourth and fifth even states } \label{Fig:density_2-5_even}
\end{figure}

\begin{figure}[h]
~\bigskip\bigskip
\par
\begin{center}
\includegraphics[width=5cm]{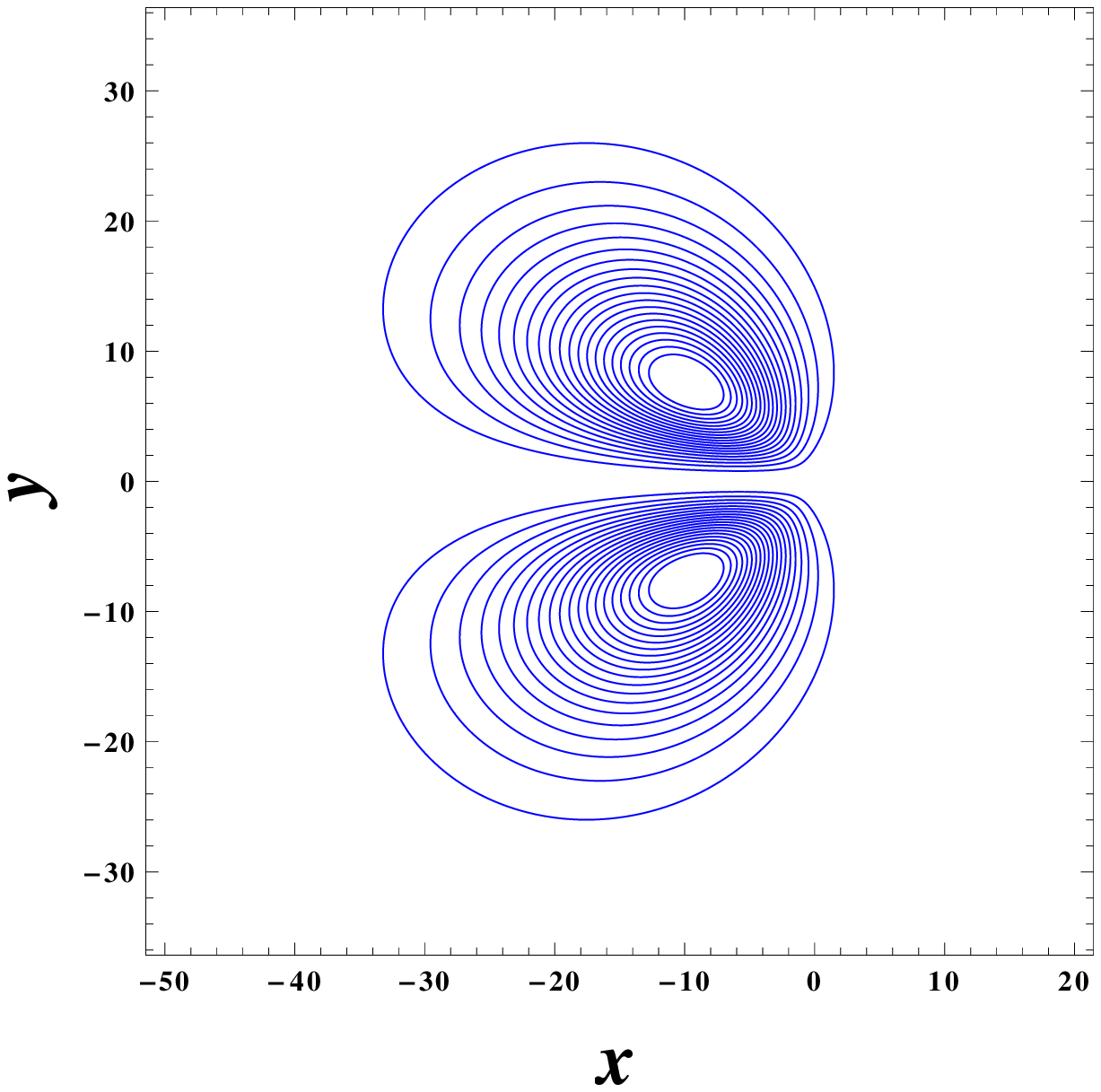} \bigskip %
\includegraphics[width=6cm]{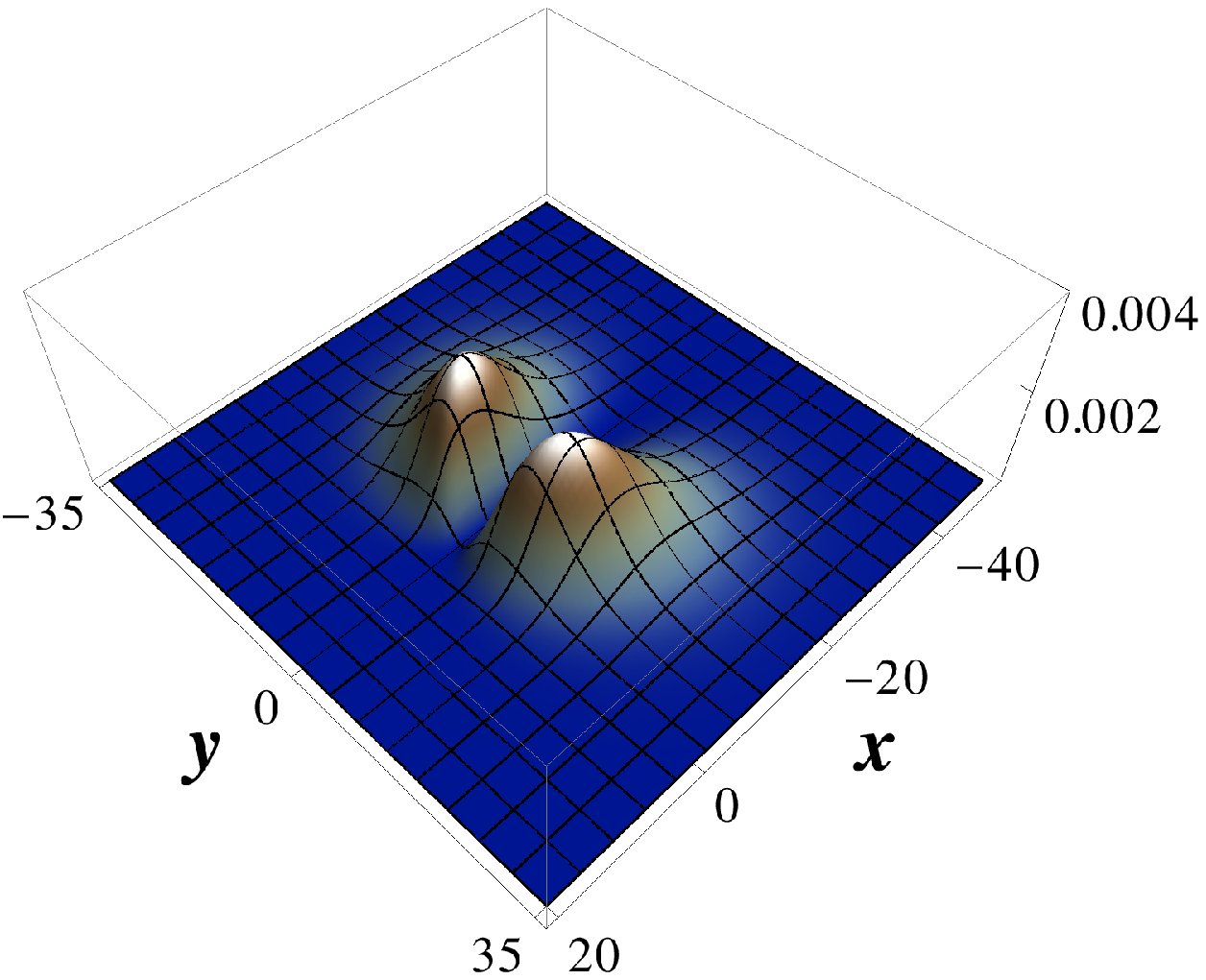} \bigskip
\end{center}
\caption{Contour and 3D plot for the square of the first odd state}
\label{Fig:density_1_odd}
\end{figure}

\begin{figure}[h]
~\bigskip\bigskip
\par
\begin{center}
\includegraphics[width=4cm]{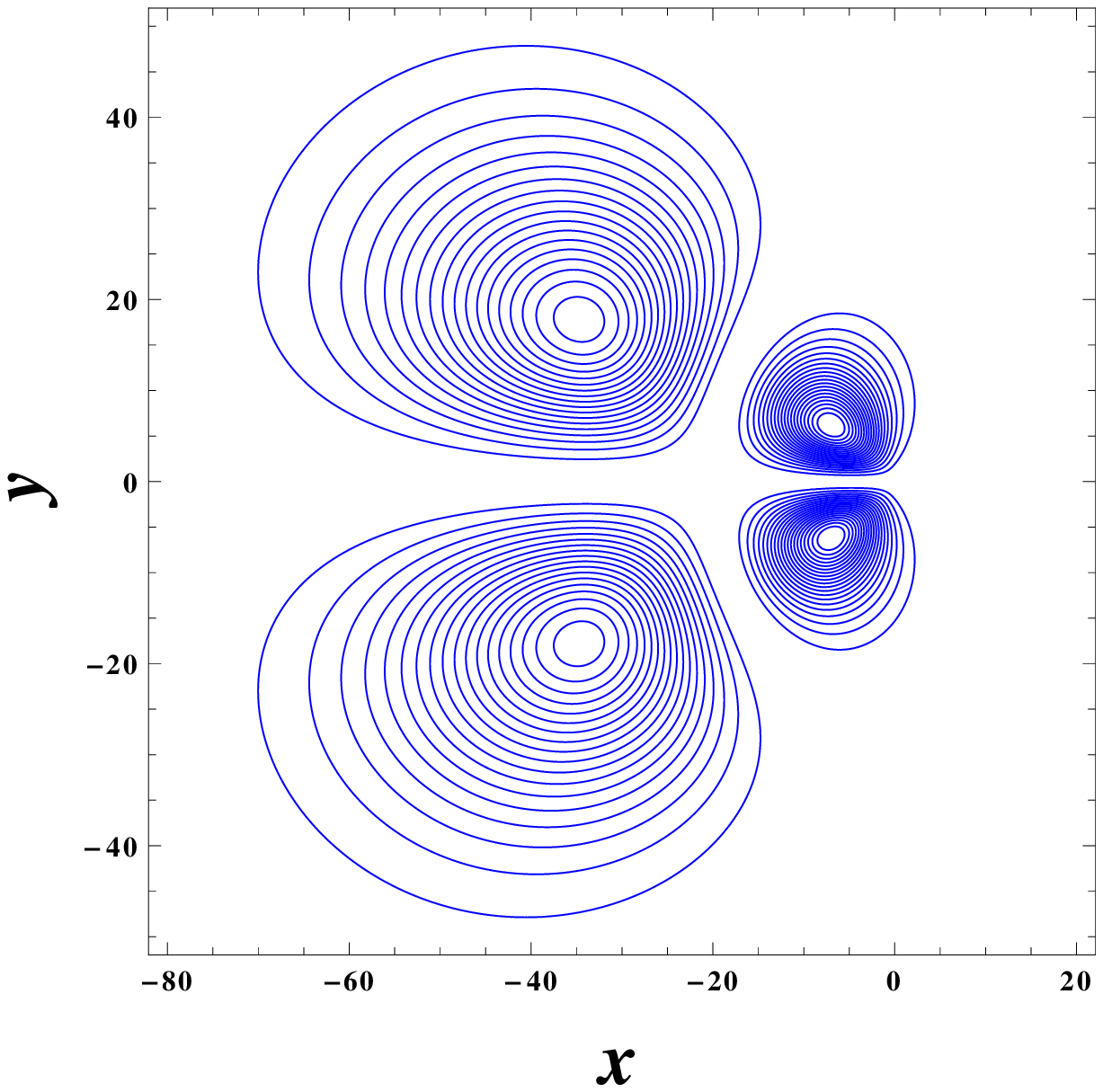} %
\includegraphics[width=4cm]{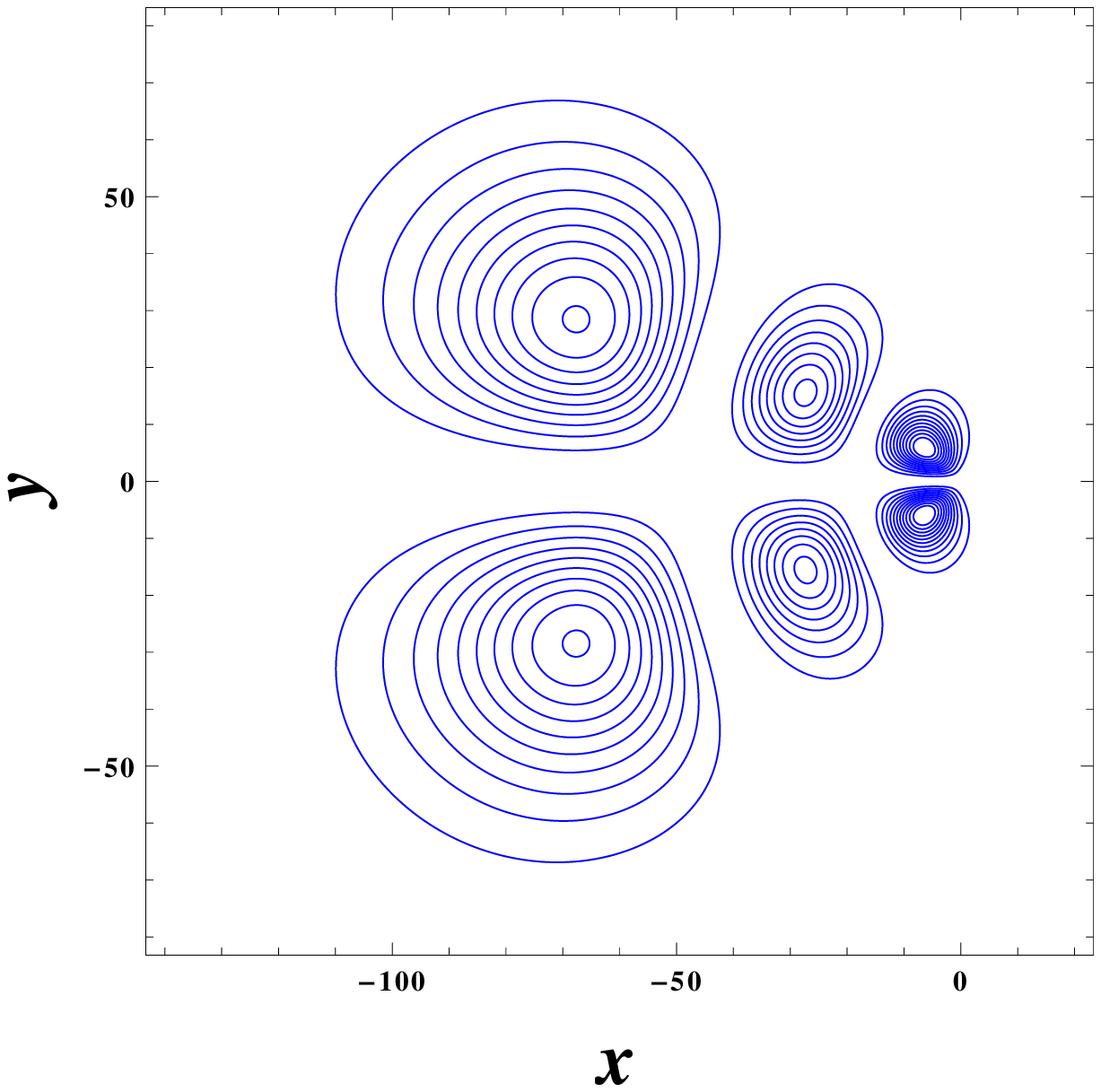} %
\includegraphics[width=4cm]{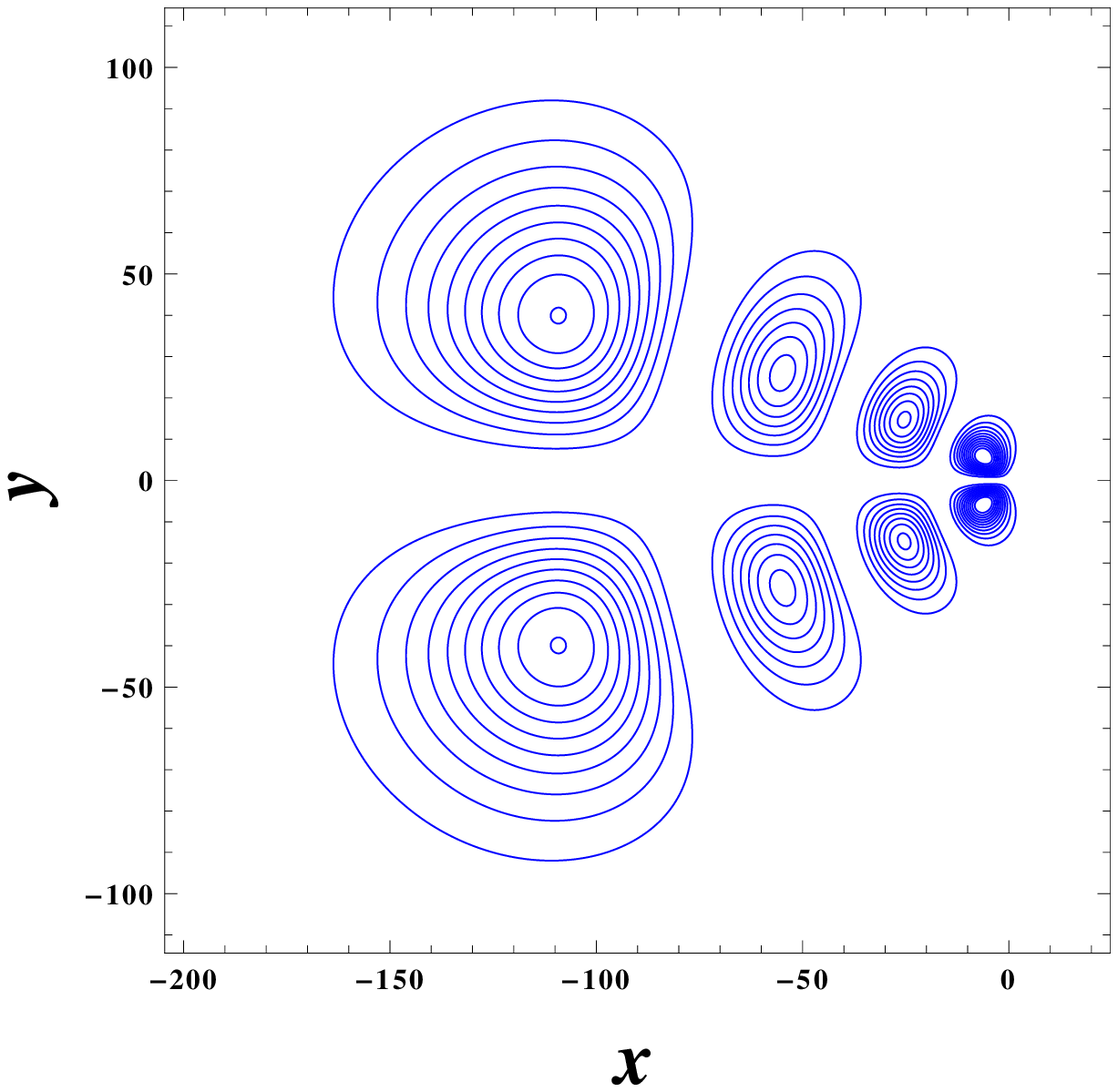} %
\includegraphics[width=4cm]{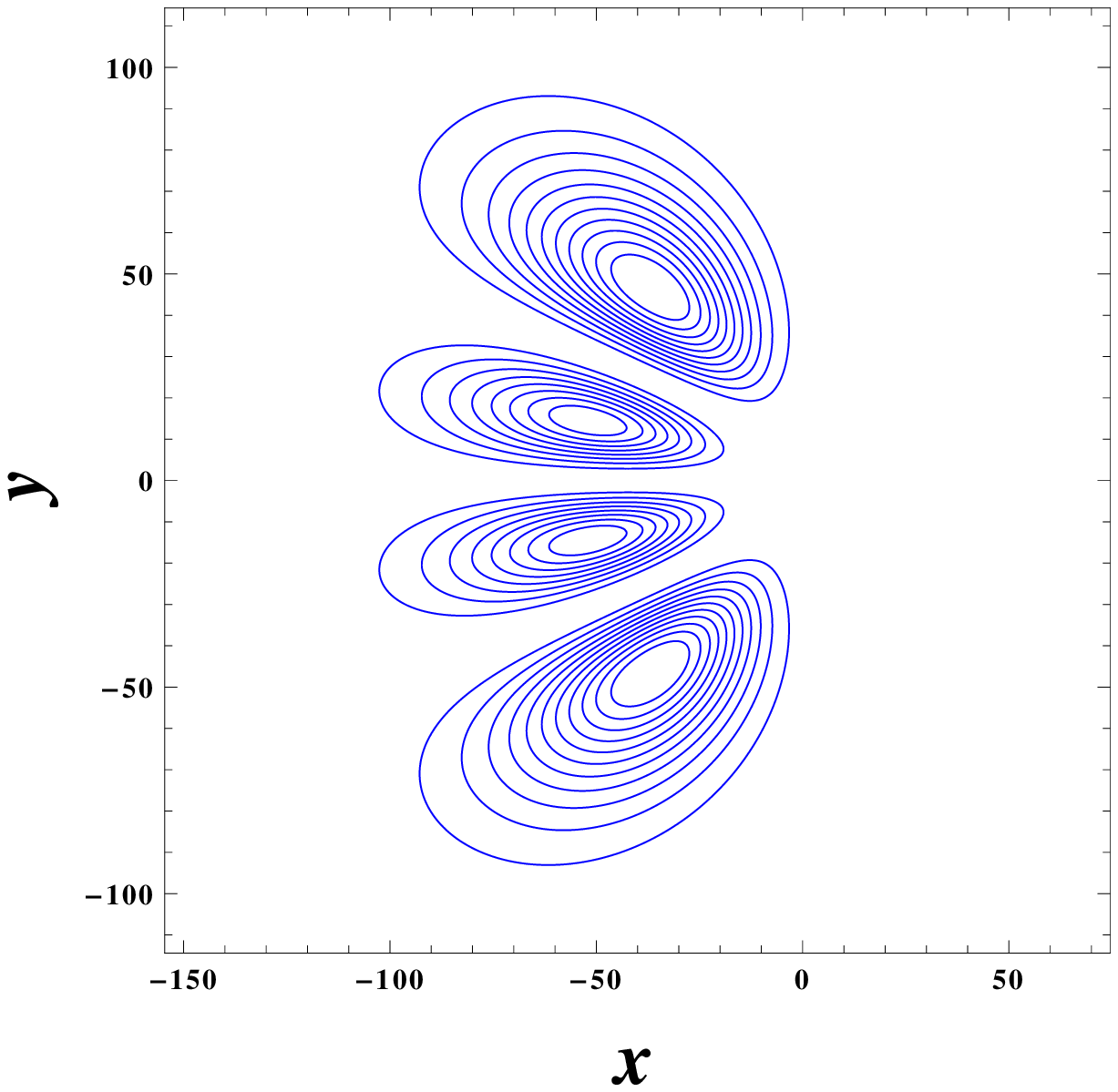} \bigskip
\end{center}
\caption{Contour plots for the squares of the second, third, fourth and
fifth odd states}
\label{Fig:density_2-5_odd}
\end{figure}

Figures \ref{Fig:density_1_even}, \ref{Fig:density_2-5_even} (left panels),
\ref{Fig:density_1_odd} and \ref{Fig:density_2-5_odd} show the contour plots
for $\psi (x,y)^{2}$ for the first five even and odd states obtained by
means of the $N=211$ NB functions. One clearly realizes that there are two
types of nodal lines $\psi (x,y)^{2}=0$ and that the energy depends
differently on each of them. The right panels of Figures \ref
{Fig:density_1_even} and \ref{Fig:density_2-5_even} show 3D plots of the
probability densities of the first even and odd states.

\section{Conclusions}

\label{sec:conclusions}

Present results clearly show that the basis set of Slater-type orbitals is
preferable to the Coulomb basis set. With just a few functions of the former
set one obtains results that are considerably more accurate than those
arising from much larger sets of the latter. As argued above, the reason is
that any linear combination of discrete-spectrum Coulomb eigenfunctions is
orthogonal to the continuous-spectrum eigenfunctions. On the other hand, no
continuous-spectrum functions are required when using the Slater-type basis
set. The contribution of the continuous spectrum appears to be more relevant
for the lowest states and for the even ones. However, we have not proved
that the RR variational method with the Slater-type functions converges
towards the actual eigenvalues as $K$ increases. Present variational
eigenvalues converge to limits that are slightly larger than the RSDM ones
and we cannot safely state that all the stable digits of our results agree
with those of the exact eigenvalues. Accurate lower bounds are required for
that purpose and we have not yet been able to obtain them. In spite of this
fact, it is encouraging that present RR results agree with the RSDM ones
within the reported $2\%$ accuracy of the latter\cite{DGYD10}. If, as argued
by Dasbiswas et al\cite{DGYD10}, the RR variational method is more
convenient than the RSDM for highly excited states, then present
contribution is relevant because there is no doubt that the basis set
proposed in this paper is preferable to the Coulomb one.

\ack P.A. acknowledges support of Conacyt through the SNI
fellowship and also of PIFI. F.M.F acknowledges support of PIFI
and of UNLP through the ``subsidio para viajes y/o estad\'{i}as''

\begin{table}[h]
\caption{Optimal $\alpha$ and energies for the first 5 even and
odd states for $K=20$} \label{tab:energies}
\begin{tabular}{c|cc|c|cc|c}
$n$ & $\alpha$ & $\epsilon_n^{NB\,even}$ & $\epsilon_n^{CB\,even}$ & $\alpha$
& $\epsilon_n^{NB\,odd}$ & $\epsilon_n^{CB\,odd}$ \\ \hline
1 & 1.667 & -0.1377416 & -0.1279886 & 0.3984 & -0.0232932 & -0.0232932 \\
2 & 0.7002 & -0.0411524 & -0.0394579 & 0.2469 & -0.0125862 & -0.0125862 \\
3 & 0.4273 & -0.0199679 & -0.0193729 & 0.1773 & -0.0079918 & -0.00799186 \\
4 & 0.2676 & -0.0118525 & -0.0115734 & 0.1239 & -0.0055643 & -0.00556435 \\
5 & 0.1515 & -0.0097472 & -0.0097472 & 0.0997 & -0.0053312 & -0.00533116
\end{tabular}
\bigskip\bigskip
\end{table}

\clearpage

\clearpage

\clearpage

\begin{thebibliography}{9}
\bibitem{DGYD10}  Dasbiswas K, Goswami D, Yoo C-D, and Dorsey A T 2010
\textit{Phys. Rev. B} \textbf{81} 064516.

\bibitem{YGWC91}  Yang X L, Guo S H, Wong K W, and Ching W Y 1991 \textit{%
Phys. Rev. A} \textbf{43} 1186.

\bibitem{A12}  Amore P 2012 \textit{Cent. Eur. J. Phys.} \textbf{10} 96.

\bibitem{CCCGA12}  Calderini D, Cavalli S, ColettiI C, Grossi G, and
Aquilanti V 2012 \textit{J. Chem. Sci.} \textbf{124} 187.

\bibitem{M33}  MacDonald J K M 1933 \textit{Phys. Rev.} \textbf{43} 830.
\end{thebibliography}
\end{document}